\documentstyle[12pt]{article}

\advance\textheight by \topskip
\newcommand{\sla}{\kern -5.4pt /}
\newcommand{\be}{\begin{equation}}
\newcommand{\ee}{\end{equation}}
\newcommand{\bea}{\begin{eqnarray}}
\newcommand{\eea}{\end{eqnarray}}
\newcommand{\beanon}{\begin{eqnarray*}}
\newcommand{\eeanon}{\end{eqnarray*}}
\newcommand{\ba}{\begin{array}}
\newcommand{\ea}{\end{array}}
\newcommand{\bi}{\begin{itemize}}
\newcommand{\ei}{\end{itemize}}
\newcommand{\ben}{\begin{enumerate}}
\newcommand{\een}{\end{enumerate}}
\newcommand{\bc}{\begin{center}}
\newcommand{\ec}{\end{center}}

\newcommand{\ol}{\overline}
\newcommand{\dotp}{\!\cdot\!}

\newcommand{\NP}[1]{{\it Nucl.\ Phys.\ }{\bf #1}}
\newcommand{\PL}[1]{{\it Phys.\ Lett.\ }{\bf #1}}

\newcommand{\PR}[1]{{\it Phys.\ Rev.\ }{\bf #1}}

\newcommand{\HPA}[1]{{\it Helv.\ Phys.\ Acta.\ }{\bf #1}}

\begin{document}
\tolerance=100000
\input feynman
\thispagestyle{empty}
\setcounter{page}{0}

\begin{flushright}
{\large DFTT 13/94}\\
{\rm February 1994\hspace*{.5 truecm}}\\
\end{flushright}

\vspace*{\fill}

\bc
{\Large \bf A new method for computing helicity amplitudes}\\[2.cm]
{\large Alessandro Ballestrero and Ezio Maina}\\[.3 cm]
{\it Dipartimento di Fisica Teorica, Universit\`a di Torino, Italy}\\
{\it and INFN, Sezione di Torino, Italy}\\
{\it v. Giuria 1, 10125 Torino, Italy.}\\
{\it e-mail: ballestrero,maina@to.infn.it}\\
\ec

\vspace*{\fill}

\begin{abstract}
{\normalsize
\noindent
An helicity formalism for perturbative calculations is presented. It is based
on the formal insertion in spinor lines of a complete set of states built up
with unphysical spinors. It is particularly convenient when massive spinors are
present. Its application to $e^+\,e^-\,\rightarrow b\,\bar b\,W^+W^-$ is
briefly discussed.}
\end{abstract}

\vspace*{\fill}
To appear in Proceedings Nonlinear Phenomena in Complex Systems.
\par
Polatsk Belarus. February 94
\newpage
\subsection*{Introduction}
In high energy collisions many particles or partons widely separated
in phase space are often produced. The calculation of  cross
sections for these processes is made difficult by the large number of
Feynman diagrams which appear in the perturbative expansion.
This is due both to the complexity of non abelian theories and to simple
combinatorics,  which  generates more and more diagrams  when  the
number $n$ of the external particles grows. Take for example a simple
graph  in which $n$ external
vector particles are attached to a fermion line. Exchanging of the
attachments results in   $n!$ graphs which  contribute  to   the
process.
\par
If one computes unpolarized cross sections for a process with a great
number ($N$)  of Feynman diagrams with the textbook method of considering
the amplitude modulus
squared  $|A|^2$ and taking the traces, one might end up
with a prohibitive number of traces to compute. In the simple case in
which there is only one fermion line per diagram, one trace has to be
evaluated for every term of  $|A|^2$, which amounts to $N(N+1)/2$ traces.
Even if  often some of them are equal, or
easy to compute, this can be a very big number.
\par
The calculation of these processes becomes  simpler if one uses
the so called helicity-amplitudes techniques. In such approach, given
an assigned helicity to external particles, one computes the
contribution of every single diagram $k$ as a complex number $a_k$,
sums over all $k$'s and takes the modulus squared. To obtain
unpolarized cross sections one simply sums the modulus squared for
the various external helicities.
\par
The use of helicity amplitude techniques in high energy physics
dates back to the works of ref.~\cite{bj}. It has been developed
by the Calkul Collaboration \cite{calkul} and by other authors~[3-8].
An improved version of the formalism of the
Calkul collaboration has been presented in \cite{ks,mana}.
\par
We describe in the following a method \cite{abem} which is in our opinion
relatively fast and easy expecially when massive particles are present.
\par
Let us consider ( Fig.~1 ) the spinor part of a  massive ( or massless)
fermion line  with $n$ insertions.

\begin{picture}(34000,8000)
\THICKLINES
\drawline\fermion[\E\REG](3000,2000)[34000]
\drawarrow[\W\ATTIP](\pmidx,\pmidy)
\global\advance \pmidx by -1000
\global\advance \pmidy by -1300
\put (\pmidx,\pmidy){\ldots \ldots}
\global\advance \pmidy by -3000
\put (\pmidx,\pmidy){Fig. 1}
\global\advance \pmidy by 6300
\put (\pmidx,\pmidy){\ldots \ldots}
\global\advance \pfrontx by 2358
\put (\pfrontx,500){$p_1$}
\THINLINES
\global\advance \pfrontx by 2358
\global\seglength=1000 \global\gaplength=250
\drawline\scalar[\N\REG](\pfrontx,\pfronty)[3]
\global\advance \pfrontx by 2858
\put (\pfrontx,500){$p_2$}
\global\seglength=1000 \global\gaplength=250
\global\advance \fermionfrontx by 10432
\drawline\scalar[\N\REG](\fermionfrontx,\fermionfronty)[3]
\global\advance \pfrontx by 2358
\put (\pfrontx,500){$p_3$}
\global\seglength=1000 \global\gaplength=250
\global\advance \fermionbackx by -4716
\drawline\scalar[\N\REG](\fermionbackx,\fermionbacky)[3]
\global\advance \pfrontx by 2358
\put (\pfrontx,500){$p_{n+1}$}
\global\seglength=1000 \global\gaplength=250
\global\advance \fermionbackx by -5716
\drawline\scalar[\N\REG](\fermionbackx,\fermionbacky)[3]
\global\advance \pfrontx by 2358
\put (\pfrontx,500){$p_{n}$}
\end{picture}
\vskip 1.5 truecm

\noindent
It has a generic expression of the following type
\be\label{1}
T^{(n)}=
\ol{U}(p_1,\lambda_1)\chi_1(p\sla_2+\mu_2)\chi_2(p\sla_3+\mu_3)\cdots
(p\sla_{n}+\mu_{n})\chi_{n}U(p_{n+1},\lambda_{n+1})
\ee
where $\lambda_1$ and $\lambda_{n+1}$ are the polarizations of the external
fermions,  $p_1$ and $p_{n+1}$  their momenta. $p_2,\ldots ,p_{n}$ and
$\mu_2,\ldots ,\mu_{n}$ are the 4-momenta and masses appearing in the fermion
propagators. $U(p,\lambda)$ ($\ol{U}(p,\lambda)$) stands for either
 $u(p,\lambda)$ ($\bar{u}(p,\lambda)$)
or  $v(p,\lambda)$ ($\bar{v}(p,\lambda)$).
The $\chi$'s are
\be\label{chis}
\chi_i^S\equiv \chi^S(c_{r_i},c_{l_i})= c_{r_i} \left(\frac{1+\gamma_5}{2}
       \right) +c_{l_i} \left(\frac{1-\gamma_5}{2}\right)
\ee
when the  insertion corresponds to a scalar (or pseudoscalar), or
\be\label{chiv}
\chi_i^V\equiv \chi^V(\eta_i,c_{r_i},c_{l_i})= \eta\sla_i \left[
c_{r_i} \left(\frac{1+\gamma_5}{2}\right) +
             c_{l_i} \left(\frac{1-\gamma_5}{2}\right) \right]
\ee
when it corresponds to a vector particle whose `polarization' is $\eta$.
Of course $\eta$ can be the polarization vector of the external particle
or the vector resulting from a complete subdiagram which is connected
in the {\it i}--th position to the fermion line.
\par
There are normally three possible ways of evaluating a spinor line
like this.
The first consists in reducing  the expression (\ref{1}) to
a trace and then perform its evaluation \cite{cr,gp}.
This reduction  is done expressing  the matrix
$U_\alpha(p_{n+1},\lambda_{n+1})\,\ol{U}_{\beta}(p_1,\lambda_1)$ as a
combination of $\gamma$ matrices in a given representation.
The second amounts to writing
explicitely, for example in the helicity representation \cite{hz},
the components of the spinors, of the $\eta\sla_i$'s and of the $p\sla_i$
and then proceed to the multiplication of the matrices and spinors.
The other way \cite{calkul,ks} consists in decomposing every $p\sla_i$ in
sums of external momenta $k\sla_i$ and use  the relation
$k\sla=\sum_{\lambda}U(k,\lambda)\ol U(k,\lambda)+M$ (with $M=+m$ if
$U=u$, $M=-m$ if $U=v$) in order to reduce everything to the computation
 of expressions of the type $\ol U(k_i,\lambda_i)\chi U(k_j,\lambda_j)$.
\par
We get a remarkable simplification with respect to the procedures sketched
above
inserting in eq.(\ref{1}), just before every $(p\sla _i+\mu_i)$,
completeness relations formed with eigenvectors of $p\sla _i$.
To do this we  must  construct  spinors $U(p,\lambda)$  which  are
defined
also for $p$ spacelike. With this method, in addition to
reducing ourselves to the computation of expressions of the type
$\ol U(p_i,\lambda_i)\chi U(p_j,\lambda_j)$, we avoid the
proliferation  of terms due to the decomposition of the  $p\sla_i$
in terms of external momenta.

\subsection*{T functions}
One can easily costruct an example of spinors defined
for any value of $p^2$   and   satisfying   Dirac   equation   and
completeness relation,
with a straightforward generalization of those introduced in ref.\cite{ks}.
One first defines spinors $w(k_0,\lambda)$ for an auxiliary
massless vector $k_0$ satisfying
\be
w(k_0,\lambda)\bar{w}(k_0,\lambda)=\frac{1+\lambda\gamma_5}{2}k\sla_0
\ee
and with their relative phase fixed by
\be
w(k_0,\lambda)=\lambda k\sla_1 w(k_0,-\lambda),
\ee
with $k_1$  a second auxiliary vector such that $k_1^2=-1$,
$k_0\dotp k_1=0$.
Spinors for a  four momentum $p$, with
$m^2=p^2$ are then obtained as:
\be\label{uvks}
u(p,\lambda)=\frac{p\sla + m}{\sqrt{2\,p\dotp k_0}}\;w(k_0,-\lambda)
\hskip 1 truecm
v(p,\lambda)=\frac{p\sla - m}{\sqrt{2\,p \dotp k_0}}\;w(k_0,-\lambda)
\ee
\be\label{uvksconj}
\bar u(p,\lambda)=\bar w(k_0,-\lambda)\;\frac{p\sla + m}{\sqrt{2\,p \dotp k_0}}
\hskip 1 truecm
\bar v(p,\lambda)=\bar w(k_0,-\lambda)\;\frac{p\sla - m}{\sqrt{2\,p \dotp k_0}}
\ee
If $p$ is spacelike,  one of the two determination of $\sqrt{p^2}$
has  to  be chosen for $m$ in the  above  formulae,  but  physical
results will not depend on this choice.

One can readily  check that with the previous definitions,
Dirac equations
\be\label{dirac}
p\sla u(p)=+m u(p) \hskip 2 truecm
p\sla v(p)=-m v(p)
\ee
\be\label{diracconj}
\bar u(p) p\sla =+m \bar u(p) \hskip 2 truecm
\bar v(p) p\sla =-m \bar v(p)
\ee
and completeness relation
\be\label{compl}
1=\sum_\lambda\frac{u(p,\lambda)\bar{u}(p,\lambda)-v(p,\lambda)
\bar{v}(p,\lambda)}{2m}
\ee
are satisfied also when $p^2\leq 0$ and $m$ is imaginary.
\par
Let us now consider the case in which there are only two
insertions in a spinor line:
\be\label{T2}
T^{(2)}(p_1;\chi_1;p_2;\chi_2;p_3)=\ol{U}(p_1,\lambda_1)\chi_1(p\sla_2+
\mu_2)\chi_2 U(p_3,\lambda_3).
\ee
One can insert in eq. (\ref{T2}), on the left of $(p\sla_2+\mu_2)$
the relation (\ref{compl}) and make use of Dirac equations to get:
\bea
T^{(2)} & = &\frac{1}{2}\ol{U}(p_1,\lambda_1)\chi_1u(p_2,\lambda_2)\times
\bar{u}(p_2,\lambda_2)\chi_2 U(p_3,\lambda_3)\times \left( 1+{\mu_2\over
m_2}\right)+\nonumber \\ \label{T2uv}
    &   &
\frac{1}{2}\ol{U}(p_1,\lambda_1)\chi_1v(p_2,\lambda_2)\times
\bar{v}(p_2,\lambda_2)\chi_2 U(p_3,\lambda_3)\times \left( 1-{\mu_2\over
m_2}\right)
\eea
This example can  be generalized to any number of insertions and shows
that the factors $(p\sla_i+\mu_i)$ can be eliminated,
reducing all fermion lines essentially to  products of $T$ functions:
\be\label{T}
T_{\lambda_1 \lambda_2}(p_1;\chi;p_2)=\ol{U}(p_1,\lambda_1)\chi U(p_2,
\lambda_2)
\ee
defined for any value of $p_1^2$ and $p_2^2$.

The $T$ functions (\ref{T}) have a simple dependence on $m_1$
and $m_2$ and as a consequence the rules for constructing  spinor lines out
of them are simple.
They can in fact, using eqs.(\ref{uvks},\ref{uvksconj}) be written as:
\be\label{Texpr}
\widetilde T_{\lambda_1 \lambda_2}(p_1;\chi;p_2)\equiv
\sqrt{p_1\dotp k_0}\;\sqrt{p_2\dotp k_0}\;
T_{\lambda_1 \lambda_2}(p_1;\chi;p_2)=\hskip 4truecm
\ee
\[
A_{\lambda_1 \lambda_2}(p_1;\chi;p_2)
+M_1B_{\lambda_1 \lambda_2}(p_1;\chi;p_2)
 +M_2C_{\lambda_1 \lambda_2}(p_1;\chi;p_2)
+M_1M_2D_{\lambda_1 \lambda_2}(p_1;\chi;p_2)\nonumber
\]
where
\bea\label{UM}
M_i=+m_i \hskip 1truecm \mbox{if}\hskip 1truecm
U(p_i,\lambda_i)=u(p_i,\lambda_i) \\
M_i=-m_i \hskip 1truecm \mbox
{if}\hskip 1truecm
U(p_i,\lambda_i)=v(p_i,\lambda_i).\nonumber
\eea
The functions $A$, $B$, $C$, $D$ turn out to be independent of
$m_1$ and $m_2$ and of the $u$ or $v$ nature of $\ol{U}(p_1,\lambda_1)$
and $U(p_2,\lambda_2)$.
We give in Appendix A the  expressions for $A^V$, $B^V$, $C^V$,
$D^V$ and $A^S$, $B^S$, $C^S$, $D^S$, which are the $A$, $B$, $C$, $D$
functions for a vector and a scalar insertion respectively .

\subsection*{From T functions to spinor lines}
The functions
\be\label{titilde}
\widetilde T^{(n)}=T^{(n)} \sqrt{p_1\dotp k_0}\;\sqrt{p_{n+1}\dotp k_0}
\; (p_2\dotp k_0) \; (p_3\dotp k_0) \cdots \; (p_n\dotp k_0).
\ee
can  be  computed  recursively starting  from  $T$  functions  (or
$\widetilde T=\widetilde T^{(1)}$).  The  $T^{(n)}$ themselves, and
hence the complete spinor line, can then be immediately obtained
at the end of the computation, dividing by the appropriate factors.
\par
Let us denote with $\widetilde T,A,B,C,D$ the $2x2$ matrices whose elements are
$\widetilde T_{\lambda_1 \lambda_2}$, $A_{\lambda_1 \lambda_2}$,
$B_{\lambda_1\lambda_2}$,
$C_{\lambda_1 \lambda_2}$, $D_{\lambda_1 \lambda_2}$.
With this notation, making use of eqs. (\ref{Texpr}) and (\ref{T}),
eq. (\ref{T2uv}) reads:
\bea
\lefteqn{ \widetilde T^{(2)}(1,2,3)  =
\frac{1}{2}\biggl[\Bigl(A(1,2)+M_1B(1,2)+m_2C(1,2)+
M_1m_2D(1,2)\Bigr)}\nonumber \\
& & \times \left( 1+{\mu_2\over m_2}\right) \times
 \Bigl(A(2,3)+m_2B(2,3)+M_3C(2,3)+m_2M_3D(2,3)\Bigr)+\label{multi}\\
& &  \Bigl(A(1,2)+M_1B(1,2)-m_2C(1,2)-M_1m_2D(1,2)\Bigr)
 \times \nonumber \\
& &  \left( 1-{\mu_2\over m_2}\right)\times
  \Bigl(A(2,3)-m_2B(2,3)+M_3C(2,3)-m_2M_3D(2,3)\Bigr)\biggr]
\nonumber
\eea
where we have used the shorthands $(1,2)$ and $(1,2,3)$ for
$(p_1;\chi_1;p_2)$ and $(p_1;\chi_1;p_2;\chi_2;p_3)$ respectively.
\par
It is immediate to see that $\widetilde T^{(2)}$ has again the same
dependence on the
external  masses as in (\ref{Texpr}):
\be\label{T2/2}
\widetilde T^{(2)}(1,2,3)=A^{(2)}(1,2,3)+M_1B^{(2)}(1,2,3)+M_3C^{(2)}(1,2,3)
+M_1M_3D^{(2)}(1,2,3)
\ee
with
\bea
A^{(2)}(1,2,3)=A(1,2)\Bigl(A(2,3)+\mu_2B(2,3)\Bigr)
  +C(1,2)\Bigl(\mu_2A(2,3)+p_2^2B(2,3)\Bigr)\nonumber\\
B^{(2)}(1,2,3)=B(1,2)\Bigl(A(2,3)+\mu_2B(2,3)\Bigr)
  +D(1,2)\Bigl(\mu_2A(2,3)+p_2^2B(2,3)\Bigr)\label{abcd}\\
C^{(2)}(1,2,3)=A(1,2)\Bigl(C(2,3)+\mu_2D(2,3)\Bigr)
  +C(1,2)\Bigl(\mu_2C(2,3)+p_2^2D(2,3)\Bigr)\nonumber\\
D^{(2)}(1,2,3)=B(1,2)\Bigl(C(2,3)+\mu_2D(2,3)\Bigr)
  +D(1,2)\Bigl(\mu_2C(2,3)+p_2^2D(2,3)\Bigr)\nonumber
\eea
This implies that $A^{(2)}$, $B^{(2)}$, $C^{(2)}$, $D^{(2)}$ can be reinserted
in an equation like eq. (\ref{multi}) to give the $\widetilde T$
function $\widetilde T^{(3)}$
corresponding to a fermion line with 3 insertions, and so on.
So one can generalize
eqs. (\ref{multi},\ref{T2/2},\ref{abcd})  by induction. Every
$\widetilde T^{(i)}$ will turn out to be of the form
\be\label{Ti}
\widetilde T^{(i)}=A^{(i)}+M_1B^{(i)}+M_{i+1}C^{(i)}+M_1M_{i+1}D^{(i)}.
\ee
{}From this it follows that the evaluation of any spinor line can be
performed
computing the $A$, $B$, $C$, $D$ matrices relative to every single insertion
and combining them toghether until one gets
to the final $T^{(n)}$.
The formalism can be convenientely cast in  matrix notation, since
every piece of a spinor line as well as every complete spinor line with $n$
insertions is completely known when we know the matrix
\be
\tau=\left(\ba{cc}A&C\\B&D\ea\right)
\ee
The law of composition  of two pieces of spinor line ,
connected by a fermion propagator with 4-momentum $p$ and mass $\mu$,
whose matrices are
\[
\tau_1=\left(\ba{cc}A_1&C_1\\B_1&D_1\ea\right)
\hskip 2truecm
\tau_2=\left(\ba{cc}A_2&C_2\\B_2&D_2\ea\right) ,
\]
is simply (cfr. eq. (\ref{abcd}) ) :
\be\label{comp}
\left(\ba{cc}A&C\\B&D\ea\right)  =
\left(\ba{cc}A_1&C_1\\B_1&D_1\ea\right)
\left(\ba{cc}1&\mu\\\mu&p^2\ea\right)
\left(\ba{cc}A_2&C_2\\B_2&D_2\ea\right).
\ee
If we call $\pi_i$ the matrix
\be\label{pi}
\pi_i=\left(\ba{cc}1&\mu_i\\\mu_i&p_i^2\ea\right)
\ee
corresponding to the propagator of 4-momentum $p_i$
and $\tau_i$ the matrix associated with the $i$-th insertion of fig.~1,
the $\tau$ matrix  of the whole spinor line can  then be computed as follows:
\be\label{tau}
\tau=\tau_1\pi_2\tau_2\pi_3\tau_3\cdots\pi_{n-1}\tau_{n-1}\pi_{n}\tau_{n}
\ee
\par
When one has to deal with massless spinor lines, all the formulae given in the
appendix A remain valid. But in this case the fact that all $\mu_i$'s as
well as $m_1$ and $m_{n+1}$ are zero leads to significant simplifications.
The $\pi_i$ matrices (\ref{pi}) become diagonal and moreover
it is not necessary to know the whole $\tau^{(n)}$ matrix to compute
the spinor line with $n$ insertions. Only $A^{(n)}$ is needed.

\subsection*{An   application:    $e^+\,e^-\,\rightarrow   b\,\bar
b\,W^+\,W^-$}
We have tested the method  for helicity  amplitude
computations  just described and we have always found perfect numerical
agreement  with   the other  methods.
We think  it   has some advantages over the  ones  mentioned  in  the
introduction and it surely turned out to be faster in our  tests.
Recently we have used it for computing  $e^+\,e^-\,\rightarrow   b\,\bar
b\,W^+\,W^-$\cite{bbww}. Some representative tree level diagrams
for this process are
reported  in Fig.~2.  From them one can see that the $W$'s can be
both attached  to  the  massive
fermion line directly  or through a trilinear
coupling  with  $\gamma$,$Z$  or Higgs.  They  can  also be both
attached  to the massless fermion line, or one to the massless
and  one  to  the massive line.  Finally one or both  of  them  can  be
attached  through a trilinear or quadrilinear vector  coupling  or
through   a   Higgs  to  the  intermediate   vector   propagators.
Altoghether the process is described   by  $61$ diagrams.
We have performed tests of gauge invariance on the amplitudes  and
we  have  also  computed the same amplitudes with  the  method  of
ref.~\cite{hz}  as a check of our results.  We found out in this  case
that the computation with our method was about four times faster.
\par
The   phenomenological  interest  of  the  reaction   above   lies
principally in the
studies  of top  and Higgs production at the Next Linear Collider. This
can  be easily understood from the first and last diagram of
fig.~2.  In  the  first  diagram a $t\bar t$  pair is  produced  with
the subsequent decay of  $t$ to $W^+b$ and $\bar t$ to $W^-\bar b$.
A  top  heavier  than $100$ GeV is expected to  decay  to  $W\,b$
before any hadronization takes place and hence, with the present
limits on the top mass, the final state under scrutiny  is precisely what
one expects to detect in  $t\,\bar t$ production.
Similarly  the  last  diagram corresponds to  $e^+\,e^-\rightarrow
H\,Z$ with subsequent decay of the Higgs in two $W$'s and of the
$Z$ in $b\,\bar b$.  The production of  $H$ in association with
$Z$ is the  most  favourable
mechanism  for  Higgs  production at an $e^+\,e^-$ collider with
an  energy  $\leq 500$ GeV and the decay channel $H\rightarrow WW$ is
the  one  with the largest branching ratio for a Higgs mass  above the $WW$
threshold.   From  the  point  of  view  of  $t\,\bar  t$  (Higgs)
production all diagrams except the first (last) in fig.~2 have  to
be considered as background.  Our amplitudes allow a complete  and
coherent  study  of  signal and background  for  these  processes,
taking into account interference and finite width effects. They also
allow the study of possible interplays between the two processes which
can be significant for particular values of the  top and
Higgs masses
. Our results concerning these processes can be found in ref.~\cite{bbww}.

\subsection*{Conclusions}
We  have described a new method for helicity calculations  based  on
the  insertion in spinor lines  of completeness  relations  formed
with  generalized spinors.  The technique has been tested  with
success and it appears to be particularly fast when massive spinors
are involved.  An
application to a reaction relevant for top and Higgs measurements
has been reported.

\subsection*
{Appendix A}

\renewcommand{\theequation}{A.\arabic{equation}}
\setcounter{equation}{0}

Using the spinors (\ref{uvks},\ref{uvksconj}, \ref{UM}),  their products can
be written as

\bea
\lefteqn{U(p_2,\lambda_2)\ol U(p_1,\lambda_1)= \frac{1}{4 \sqrt{2\,p_1\dotp
k_0}
\sqrt{2\, p_2\dotp k_0} }}\label{uubar}\\
& & \times(p\sla_2+M_2)\left[ (1+\lambda_1\lambda_2)-
(\lambda_1+\lambda_2)\gamma^5+k\sla_1[(\lambda_1-\lambda_2)-
(1-\lambda_1\lambda_2)\gamma^5]\right]k\sla_0(p\sla_1+M_1).\nonumber
\eea
Multiplying to the right by $\chi ^S$  or $\chi ^V$ of
eqs.~(\ref{chis},\ref{chiv}) , taking the trace  and with the help of
eq.~(\ref {Texpr}), one gets the following expressions for the functions
$A$, $B$, $C$, $D$ corresponding to a scalar and a vector insertion :

\bea\label{a2}
A^S_{+-}&=&c_l\left(k_0\dotp p_1\;k_1\dotp p_2-
          k_0\dotp p_2\;k_1\dotp p_1-i
\epsilon\,(k_0,k_1,p_1,p_2)\right)\nonumber\\
A^S_{-+}&=&c_r\left(-k_0\dotp p_1\;k_1\dotp p_2+
          k_0\dotp p_2\;k_1\dotp p_1-i
\epsilon\,(k_0,k_1,p_1,p_2)\right)\nonumber\\
B^S_{++}&=&c_r\; k_0\dotp p_2 \hskip 2cm
B^S_{--}\;=\;c_l\; k_0\dotp p_2\nonumber\\
C^S_{++}&=&c_l\; k_0\dotp p_1\hskip 2cm
C^S_{--}\;=\;c_r\; k_0\dotp p_1\nonumber\\
A^V_{++}&=&c_r\left(-k_0\dotp \eta\;p_1\dotp p_2+k_0\dotp p_1\;\eta\dotp p_2+
    k_0\dotp p_2\;\eta\dotp p_1+ i
\epsilon\,(k_0,\eta,p_1,p_2)\right)\nonumber\\
A^V_{--}&=&c_l\left(-k_0\dotp \eta\;p_1\dotp p_2+k_0\dotp p_1\;\eta\dotp p_2+
    k_0\dotp p_2\;\eta\dotp p_1- i \epsilon\,(k_0,\eta,p_1,p_2)\right)\\
B^V_{+-}&=&c_l\left(k_0\dotp \eta\; k_1\dotp p_2
            - k_0\dotp p_2 \; k_1\dotp \eta
         -i \epsilon\,(k_0,k_1,\eta,p_2)\right)\nonumber\\
B^V_{-+}&=&c_r\left(
        -k_0\dotp \eta\; k_1\dotp p_2 + k_0\dotp p_2 \; k_1\dotp \eta
        -i \epsilon\,(k_0,k_1,\eta,p_2)\right)\nonumber\\
C^V_{+-}&=&c_r\left(
        -k_0\dotp \eta\; k_1\dotp p_1 + k_0\dotp p_1\; k_1\dotp \eta
           +i \epsilon\,(k_0,k_1,\eta,p_1)\right)\nonumber\\
C^V_{-+}&=&c_l \left(
        k_0\dotp \eta\; k_1\dotp p_1 - k_0\dotp p_1\; k_1\dotp \eta
           +i \epsilon\,(k_0,k_1,\eta,p_1)\right)\nonumber\\
D^V_{++}&=&c_l\; k_0\dotp\eta\hskip 2cm
D^V_{--}\;=\;c_r\; k_0\dotp\eta.\nonumber
\eea
All functions $A$, $B$, $C$, $D$ for a single insertion not reported in the
preceding list are identically zero.
\par\noindent
The function $\epsilon$ is defined to be the determinant:
\be
\epsilon\,(p,q,r,s)= det \;
\left|\ba{cccc}       p^0 & q^0 & r^0 & s^0\\
                      p^1 & q^1 & r^1 & s^1\\
                      p^2 & q^2 & r^2 & s^2\\
                      p^3 & q^3 & r^3 & s^3 \ea\right|
\ee

\newpage
\
\vskip 2.0cm

\begin{picture}(10000,8000)
\THICKLINES
\bigphotons
\drawline\photon[\W\REG](10000,8000)[6]
\drawline\fermion[\NW\REG](\photonbackx,\photonbacky)[5000]
\drawarrow[\SE\ATBASE](\pmidx,\pmidy)
\drawline\fermion[\SW\REG](\photonbackx,\photonbacky)[5000]
\drawarrow[\SW\ATBASE](\pmidx,\pmidy)
\drawline\fermion[\NE\REG](\photonfrontx,\photonfronty)[5000]
\drawarrow[\NE\ATBASE](\pmidx,\pmidy)
\drawline\photon[\E\REG](\pmidx,\pmidy)[4]
\drawline\fermion[\SE\REG](10000,8000)[5000]
\drawarrow[\NW\ATBASE](\pmidx,\pmidy)
\drawline\photon[\E\REG](\pmidx,\pmidy)[4]
\put(-500,12000){$e^-$}
\put(-500,3000){$e^+$}
\put(14000,12000){$b$}
\put(14000,3000){$\bar b$}
\put(16250,9250){$W^+$}
\put(16250,5750){$W^-$}
\put(6500,2000){$(1)$}
\drawline\photon[\W\REG](32000,8000)[6]
\drawline\fermion[\NW\REG](\photonbackx,\photonbacky)[5000]
\drawarrow[\SE\ATBASE](\pmidx,\pmidy)
\drawline\fermion[\SW\REG](\photonbackx,\photonbacky)[5000]
\drawarrow[\SW\ATBASE](\pmidx,\pmidy)
\drawline\fermion[\NE\REG](\photonfrontx,\photonfronty)[5000]
\drawarrow[\NE\ATBASE](\pmidx,\pmidy)
\drawvertex\photon[\E 3](\pmidx,\pmidy)[3]
\drawline\fermion[\SE\REG](32000,8000)[5000]
\drawarrow[\NW\ATBASE](\pmidx,\pmidy)
\put(21500,12000){$e^-$}
\put(21500,3000){$e^+$}
\put(36000,12000){$b$}
\put(36000,3000){$\bar b$}
\put(39300,11700){$W^+$}
\put(39300,7050){$W^-$}
\put(28500,2000){$(2)$}
\end{picture}

\vskip 2.0cm

\begin{picture}(10000,8000)
\THICKLINES
\bigphotons
\drawline\photon[\W\REG](10000,8000)[6]
\drawline\fermion[\NW\REG](\photonbackx,\photonbacky)[5000]
\drawarrow[\SE\ATBASE](\pmidx,\pmidy)
\drawline\photon[\E\REG](\pmidx,\pmidy)[4]
\drawline\fermion[\SW\REG](4000,8000)[5000]
\drawarrow[\SW\ATBASE](\pmidx,\pmidy)
\drawline\photon[\E\REG](\pmidx,\pmidy)[4]
\drawline\fermion[\NE\REG](10000,8000)[5000]
\drawarrow[\NE\ATBASE](\pmidx,\pmidy)
\drawline\fermion[\SE\REG](10000,8000)[5000]
\drawarrow[\NW\ATBASE](\pmidx,\pmidy)
\put(-500,12000){$e^-$}
\put(-500,3000){$e^+$}
\put(14000,12000){$b$}
\put(14000,3000){$\bar b$}
\put(6500,9500){$W^-$}
\put(6500,5500){$W^+$}
\put(6500,2000){$(3)$}
\drawline\photon[\W\REG](32000,8000)[6]
\drawline\fermion[\NW\REG](\photonbackx,\photonbacky)[5000]
\drawarrow[\SE\ATBASE](\pmidx,\pmidy)
\drawline\photon[\E\REG](\pmidx,\pmidy)[4]
\drawline\fermion[\SW\REG](26000,8000)[5000]
\drawarrow[\SW\ATBASE](\pmidx,\pmidy)
\drawline\fermion[\NE\REG](32000,8000)[5000]
\drawarrow[\NE\ATBASE](\pmidx,\pmidy)
\drawline\photon[\E\REG](\pmidx,\pmidy)[4]
\drawline\fermion[\SE\REG](32000,8000)[5000]
\drawarrow[\NW\ATBASE](\pmidx,\pmidy)
\put(21500,12000){$e^-$}
\put(21500,3000){$e^+$}
\put(36000,12000){$b$}
\put(36000,3000){$\bar b$}
\put(38000,9500){$W^+$}
\put(28500,9500){$W^-$}
\put(28500,2000){$(4)$}
\end{picture}

\vskip 2.0cm

\begin{picture}(10000,8000)
\THICKLINES
\bigphotons
\drawline\photon[\W\REG](9000,8000)[5]
\drawline\fermion[\NW\REG](\photonbackx,\photonbacky)[5000]
\drawarrow[\SE\ATBASE](\pmidx,\pmidy)
\drawline\photon[\E\REG](\pmidx,\pmidy)[3]
\drawline\fermion[\SW\REG](\fermionfrontx,\fermionfronty)[5000]
\drawarrow[\SW\ATBASE](\pmidx,\pmidy)
\drawline\photon[\NE\REG](9000,8000)[6]
\seglength=1416  \gaplength=300  
\drawline\scalar[\SE\REG](\photonfrontx,\photonfronty)[2]
\drawline\fermion[\NE\REG](\scalarbackx,\scalarbacky)[3000]
\drawarrow[\NE\ATBASE](\pmidx,\pmidy)
\drawline\fermion[\SE\REG](\scalarbackx,\scalarbacky)[3000]
\drawarrow[\NW\ATBASE](\pmidx,\pmidy)
\put(-500,12000){$e^-$}
\put(-500,3000){$e^+$}
\put(13000,12000){$W^+$}
\put(5650,9500){$W^-$}
\put(13850,8000){$b$}
\put(9000,5500){$H$}
\put(13850,2750){$\bar b$}
\put(6000,2000){$(5)$}
\drawline\photon[\W\REG](31000,8000)[5]
\drawline\fermion[\NW\REG](\photonbackx,\photonbacky)[5000]
\drawarrow[\SE\ATBASE](\pmidx,\pmidy)
\drawline\fermion[\SW\REG](\photonbackx,\photonbacky)[5000]
\drawarrow[\SW\ATBASE](\pmidx,\pmidy)
\drawline\photon[\NE\REG](\photonfrontx,\photonfronty)[6]
\drawline\photon[\SE\REG](31000,8000)[4]
\drawline\fermion[\NE\REG](\photonbackx,\photonbacky)[3000]
\drawarrow[\NE\ATBASE](\pmidx,\pmidy)
\drawline\fermion[\SE\REG](\photonbackx,\photonbacky)[3000]
\drawarrow[\NW\ATBASE](\pmidx,\pmidy)
\drawline\photon[\E\REG](\pmidx,\pmidy)[3]
\put(21500,12000){$e^-$}
\put(21500,3000){$e^+$}
\put(35000,12000){$W^+$}
\put(37900,4200){$W^-$}
\put(35750,8000){$b$}
\put(35750,2250){$\bar b$}
\put(28000,2000){$(6)$}
\end{picture}

\vskip 2.0cm

\begin{picture}(10000,8000)
\THICKLINES
\bigphotons
\drawline\photon[\W\REG](10000,8000)[6]
\drawline\fermion[\NW\REG](\photonbackx,\photonbacky)[5000]
\drawarrow[\SE\ATBASE](\pmidx,\pmidy)
\drawline\fermion[\SW\REG](\photonbackx,\photonbacky)[5000]
\drawarrow[\SW\ATBASE](\pmidx,\pmidy)
\drawline\photon[\NE\REG](\photonfrontx,\photonfronty)[6]
\drawline\photon[\E\REG](12300,10000)[3]
\drawline\fermion[\NE\REG](\photonbackx,\photonbacky)[3000]
\drawarrow[\NE\ATBASE](\pmidx,\pmidy)
\drawline\fermion[\SE\REG](\photonbackx,\photonbacky)[3000]
\drawarrow[\NW\ATBASE](\pmidx,\pmidy)
\drawline\photon[\SE\REG](10000,8000)[6]
\put(-500,12000){$e^-$}
\put(-500,3000){$e^+$}
\put(14000,12000){$W^+$}
\put(14000,3000){$W^-$}
\put(17650,12000){$b$}
\put(17650,6750){$\bar b$}
\put(6000,2000){$(7)$}
\drawline\photon[\W\REG](31000,8000)[5]
\drawline\fermion[\NW\REG](\photonbackx,\photonbacky)[5000]
\drawarrow[\SE\ATBASE](\pmidx,\pmidy)
\drawline\fermion[\SW\REG](\photonbackx,\photonbacky)[5000]
\drawarrow[\SW\ATBASE](\pmidx,\pmidy)
\seglength=1416  \gaplength=300  
\drawline\scalar[\NE\REG](\photonfrontx,\photonfronty)[3]
\drawline\photon[\NE\REG](\scalarbackx,\scalarbacky)[4]
\drawline\photon[\SE\REG](\scalarbackx,\scalarbacky)[4]
\drawline\photon[\SE\REG](31000,8000)[4]
\drawline\fermion[\NE\REG](\photonbackx,\photonbacky)[3000]
\drawarrow[\NE\ATBASE](\pmidx,\pmidy)
\drawline\fermion[\SE\REG](\photonbackx,\photonbacky)[3000]
\drawarrow[\NW\ATBASE](\pmidx,\pmidy)
\put(21500,12000){$e^-$}
\put(21500,3000){$e^+$}
\put(37250,13250){$W^+$}
\put(37250,8500){$W^-$}
\put(32000,10500){$H$}
\put(36000,8000){$b$}
\put(36000,2250){$\bar b$}
\put(28000,2000){$(8)$}
\end{picture}

\vskip 1.5cm
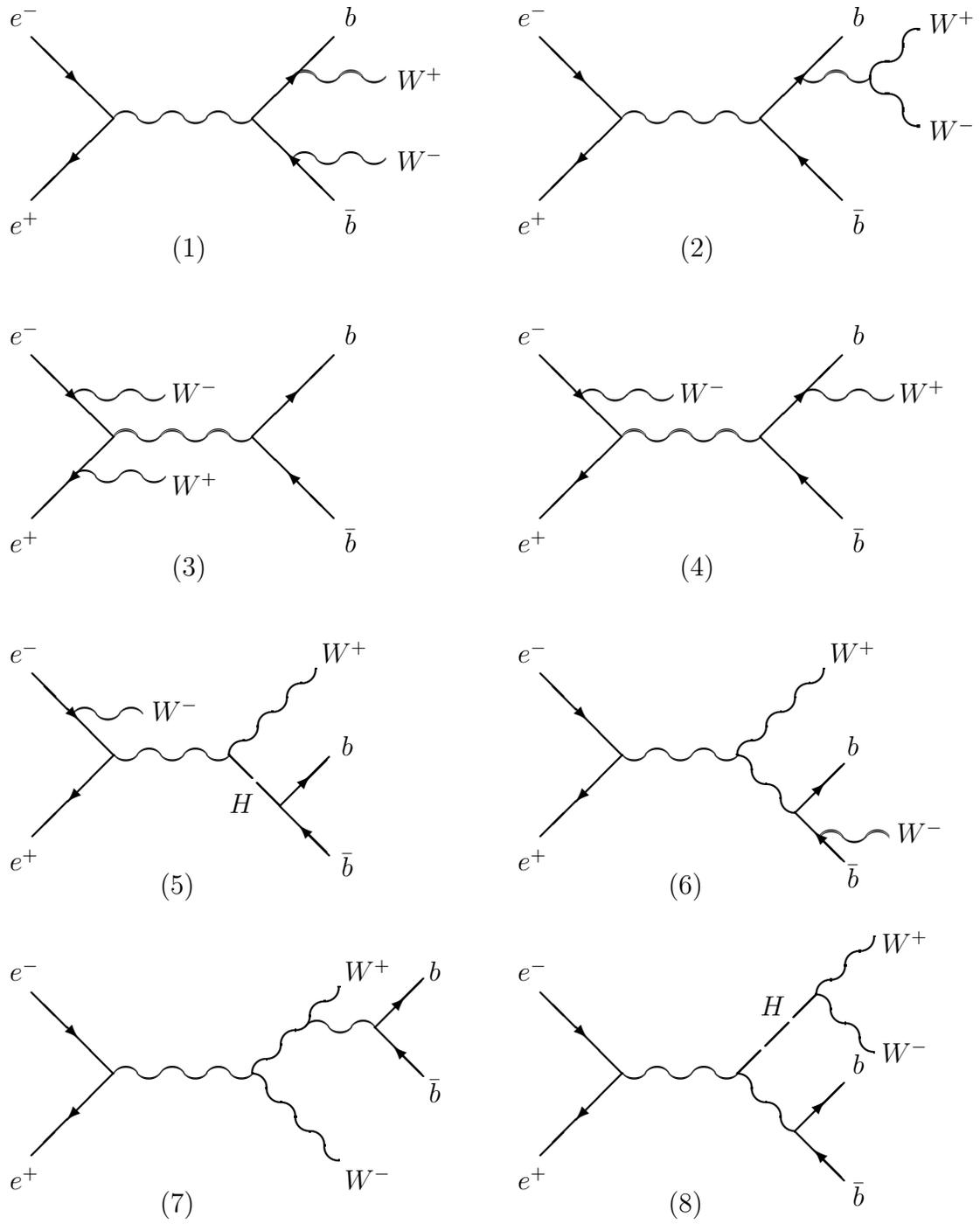
\centerline{ Fig. 2}
\vfill

\end{document}